\documentclass[aip,amsmath,amssymb,reprint]{revtex4-1}
\usepackage{graphicx}% Include figure files
\usepackage{dcolumn}% Align table columns on decimal point
\usepackage{bm}% bold math
\usepackage[utf8]{inputenc}
\usepackage[T1]{fontenc}
\usepackage{mathptmx}
\usepackage{etoolbox}
\usepackage{hyperref}

\newcommand{\Rstart}{R_{\mathrm{start}}}
\newcommand{\tRstart}{\tilde{R}_{\mathrm{start}}}
\newcommand{\Rend}{R_{\mathrm{end}}}
\newcommand{\tRend}{\tilde{R}_{\mathrm{end}}}
\newcommand{\Nturns}{N_{\mathrm{turns}}}

\draft % marks overfull lines with a black rule on the right

\begin{document}

% Use the \preprint command to place your local institutional report number 
% on the title page in preprint mode.
% Multiple \preprint commands are allowed.
%\preprint{}

\title{Utilizing bifurcations to separate particles in spiral inertial microfluidics}%

% repeat the \author .. \affiliation  etc. as needed
% \email, \thanks, \homepage, \altaffiliation all apply to the current author.
% Explanatory text should go in the []'s, 
% actual e-mail address or url should go in the {}'s for \email and \homepage.
% Please use the appropriate macro for the type of information

% \affiliation command applies to all authors since the last \affiliation command. 
% The \affiliation command should follow the other information.

\author{Rahil N. Valani}\email{rahil.valani@adelaide.edu.au}
\affiliation{School of Mathematical Sciences, University of Adelaide, South Australia 5005, Australia}
\author{Brendan Harding}\email{brendan.harding@vuw.ac.nz}
\affiliation{School of Mathematics and Statistics, Victoria University of Wellington, Wellington 6012, New Zealand}
\author{Yvonne M. Stokes}\email{yvonne.stokes@adelaide.edu.au}
\affiliation{School of Mathematical Sciences, University of Adelaide, South Australia 5005, Australia}
%\email[]{Your e-mail address}
%\homepage[]{Your web page}
%\thanks{}
%\altaffiliation{}

% Collaboration name, if desired (requires use of superscriptaddress option in \documentclass). 
% \noaffiliation is required (may also be used with the \author command).
%\collaboration{}
%\noaffiliation

\date{\today}

\begin{abstract}

Particles suspended in fluid flow through a closed duct can focus to specific stable locations in the duct cross-section due to hydrodynamic forces arising from the inertia of the disturbed fluid. Such particle focusing is exploited in biomedical and industrial technologies to separate particles by size. In curved ducts, the particle focusing is a result of balance between two dominant forces on the particle: (i) inertial lift arising from small inertia of the fluid, and (ii) drag arising from cross-sectional vortices induced by the centrifugal force on the fluid. Bifurcations of particle equilibria take place as the bend radius of the curved duct varies. By using the mathematical model of \citet{harding_stokes_bertozzi_2019}, we illustrate via numerical simulations that these bifurcations can be leveraged in a spiral duct to achieve large separation between different sized particles by transiently focusing smaller particles near saddle-points. We demonstrate this by separating similar-sized particles, as well as particles that have a large difference in size, using spiral ducts with square cross-section. The formalism of using bifurcations to manipulate particle focusing can be applied more broadly to different geometries in inertial microfluidics which may open new avenues in particle separation techniques.

\end{abstract}

\pacs{}% insert suggested PACS numbers in braces on next line

\maketitle %\maketitle must follow title, authors, abstract and \pacs

    The ability to separate particles of different sizes suspended in a fluid is important in many biomedical and industrial technologies~\cite{Liu2019}. For example, efficient isolation of rare circulating tumor cells from a large concentration of red blood cells and white blood cells in a blood sample can revolutionize cancer diagnostics and help in determining a likely prognosis~\cite{C0LC90100H}. Another example is the detection and separation of waterborne pathogens in drinking water~\cite{Waterpathogens,JIMENEZ2017247,Seo2007}. Microfluidics has become an important tool for particle separation due to small sample consumption, fast processing time, high spatial resolution and high portability~\cite{RevModPhys.77.977}. Amongst the different possible microfluidic technologies aimed at particle separation, inertial microfluidics has been used widely because of its ease of operation and high separation resolution~\cite{review1,review2}. 
    
    Segr{\'e} and Silberberg~\cite{SEGRE1961} first reported that particles suspended in fluid flow through a straight pipe with a circular cross-section can migrate across streamlines and accumulate to an annular region at approximately $60\%$ of the pipe radius. This deviation of particles from fluid streamlines is due to the inertial lift force acting on the particle that arises from small but non-negligible inertia of the disturbed fluid flow at low to moderate Reynolds numbers. This results in the phenomenon of inertial migration and subsequent focusing of particles. In straight ducts, the locations where no net force acts on the particle in the duct cross-section, henceforth referred to as {particle equilibria}, vary with the geometry of the duct cross-section~\cite{C5LC01100K}, particle size~\cite{harding_stokes_bertozzi_2019} and the flow Reynolds number~\cite{chun2006,miura_itano_sugihara-seki_2014}. Adding curvature to the duct also influences particle equilibria via the introduction of cross-sectional vortices to the flow, known as Dean vortices~\citep{Dean1927,harding_stokes_bertozzi_2019}. Curved microchannels with circular and spiral geometries are commonly used in inertial microfluidic devices aimed at particle separation by size~\citep{Liu2019}. In these channels, the interplay between (i) the inertial lift force arising from fluid inertia, and (ii) the secondary drag force arising from Dean vortices, determines the number, nature and location of particle equilibria. By tuning the relative strength of these forces via changes in the bend radius, the particle equilibria can be manipulated to achieve particle separation for different sized particles.

    %\item Give details and examples of the different methods (refere to the book chapter for more details in this) used in spiral channels and also the different cross-section and other things that are realized in spiral channels. While I am describing this, also use words like particle equilibria, limit cycle etc so that they are introduced and the reader becomes familiar with them.
    
Spiral channels with rectangular and trapezoidal cross-sections have shown promise for efficient size-based particle separation~\citep{Liu2019}. In these geometries, the various parameters are chosen such that one obtains a horizontal separation between the particle equilibria of the chosen particle sizes to be separated. The particles focus to their respective stable equilibrium points or a stable limit cycle towards the end of the spiral channel after which the channel is split into multiple channels to collect the separated particles~\citep{Rect1,Rect2,trap1,trap2}. Another separation method commonly used with circular and spiral channels having rectangular cross-section is known as Dean Flow Fractionation (DFF)~\cite{Bhagat2008,Hou2013}. In this method, the spiral channel has two inlets, one consisting of the sample containing particles (typically a mix of two distinct sizes) and the other through which a sheath flow is introduced. As the particles flow through the curved channel, the Dean vortices transport the smaller particles towards the outer wall, while a balance of inertial lift and secondary drag equilibrates the larger particles near the inner wall, thus achieving separation between the two particle sizes. Note that with this method, the smaller particles do not reach a stable equilibrium position, rather, their well-controlled migration due to Dean vortices drives their separation from the larger particles. A variant of DFF has also been developed to separate several different smaller (sub-micron) sized particles and is called High-Resolution Dean Flow Fractionation (HiDFF)~\cite{Tay2017}.     
    
Many advances in particle separation methods are primarily driven by experimental trial-and-error, with the potential of predicting and optimizing particle separation based on theoretical models and numerical simulations not yet being fully exploited. 
%Although the use of theoretical and numerical methods has progressed our understanding of particle equilibria and their bifurcations in straight channels~\cite{hood_lee_roper_2015,Fox_2020,Yamashita2019,fox_schneider_khair_2021}, only recently, progress has also been made to gain a systematic understanding of the particle equilibria in curved channels by understanding how various system parameters, such as particle size, bend radius and aspect ratio of the cross-section, can affect the location and nature of the particle equilibria~\cite{harding_stokes_bertozzi_2019,Kyung2021,ValaniDSTA2021,Valani2022SIADS}. 
Although the use of theoretical and numerical methods has progressed our understanding of particle equilibria and their bifurcations in straight channels~\cite{hood_lee_roper_2015,Fox_2020,Yamashita2019,fox_schneider_khair_2021}, only recently, progress has also been made to gain a systematic understanding of the particle equilibria in curved channels~\cite{harding_stokes_bertozzi_2019}. 
This has improved the understanding of how various system parameters, such as particle size, bend radius and aspect ratio of the cross-section, can affect the location and nature of the particle equilibria.
Subsequently, rich bifurcations in particle equilibria have been observed with respect to variations in the bend radius of the curved duct~\citep{Kyung2021,ValaniDSTA2021,Valani2022SIADS}. 
%Herein, we utilize these bifurcations in a spiral microchannel to produce large separation between two sets of different sized particles. 
Herein, we illustrate how these bifurcations can be utilized to design spiral microchannels that produce large separation between two sets of different sized particles. 
%In particular, we show that
%We show that if the size difference between the two particles is small, then one can separate the particles independent of their starting position in the cross-section. Conversely, if the size difference between the two particles is large, the desired separation is achieved by injecting the particles on the inner half of the duct cross-section. 
Although we restrict the present study to a square cross-section and 
certain particle sizes to illustrate the separation mechanism, the general formalism is applicable to a broad range of geometries in which similar bifurcations in the particle equilibria take place.

%\end{itemize}

%\section{Theoretical Formulation}

%\begin{itemize}
%    \item Start by explaining the theoretical setup of the problem with the particle flowing in a curved duct
%    \item Briefly state the equation of motion governing the dynamics of the fluid and the particle
%    \item Briefly comment on how we are solving the resulting equations of motion to obtain the particle dynamics
%    \item Also comment on the numerical method used
%    \item Refer the reader to Brendan's paper and/or Supplemental material where I can provide more details about the theoretical model and how we are solving it
%\end{itemize}

\begin{figure}
\centering
\includegraphics[width=\columnwidth]{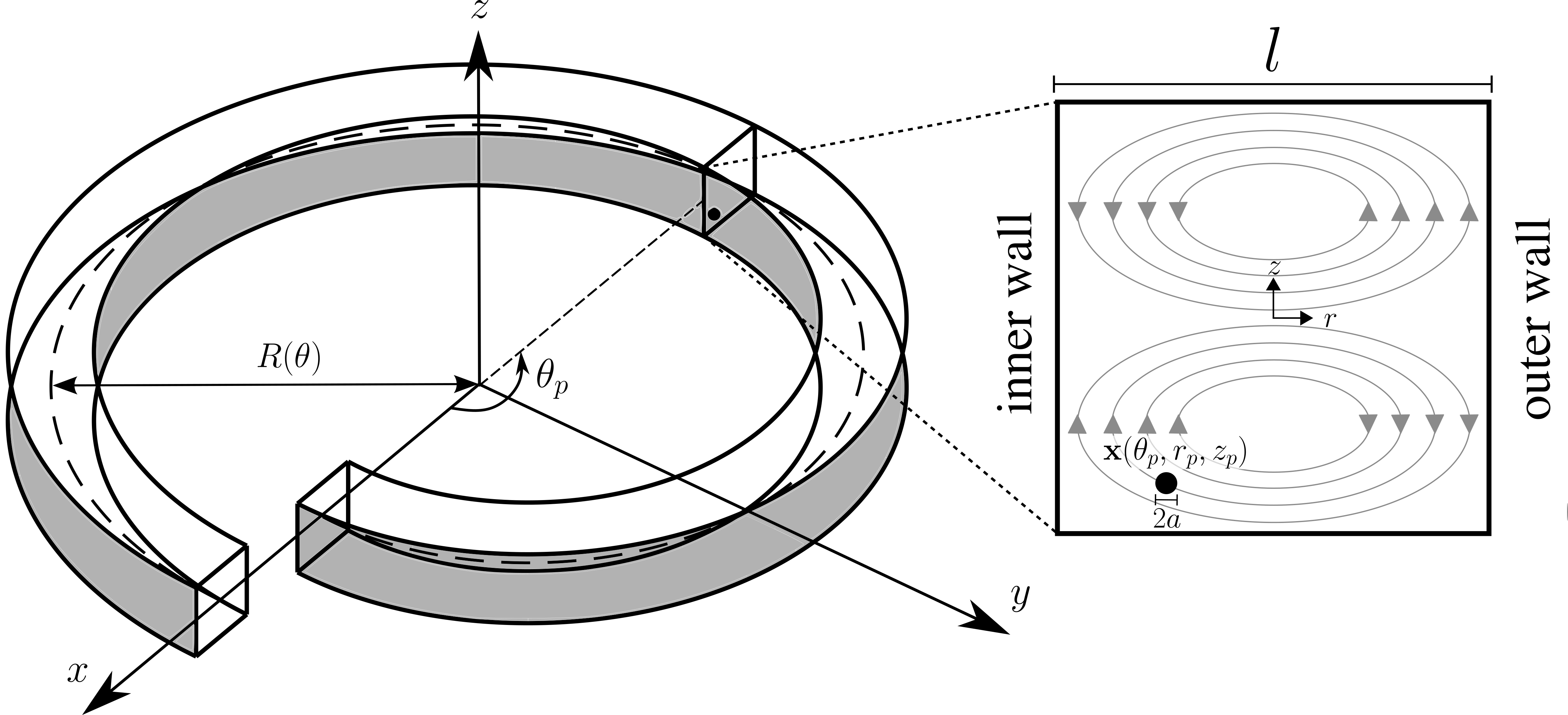}
\caption{Schematic of the theoretical setup. A particle of radius $a$ with center located at $\mathbf{x}_p=\mathbf{x}(\theta_p,r_p,z_p)$ is suspended in an incompressible fluid flow through an Archimedean spiral duct having a uniform square cross section of side length $l$. %The radius of the spiral is $R(\theta)=\Rstart+(\Delta R/2\pi \Nturns)\theta$ where $\Delta R = \Rend-\Rstart$ and $\Rstart$ and $\Rend$ are the radii of the spiral at the start and the end of the duct respectively. 
The enlarged view of the cross-section illustrates the local cross-sectional $(r,z)$ co-ordinate system, and the secondary flow (gray closed curves) induced by the curvature of the duct. The edge labeled ``inner wall" is the side closer to origin $(x,y,z)=(0,0,0)$ while the edge labeled ``outer wall" is the side further away from the origin.}
\label{Fig: Schematic}
\end{figure}

With reference to Fig.~\ref{Fig: Schematic}, consider a particle of density $\rho$ and radius $a$ suspended in an incompressible fluid flow of the same density $\rho$ and dynamic viscosity $\mu$ flowing through an Archimedean spiral duct. The instantaneous radius of the spiral varies with the azimuthal angle $\theta$ according to $R(\theta)=\Rstart+(\Delta R/2\pi \Nturns)\theta$ where $\Delta R = \Rend-\Rstart$ is the change in radius from the start to the end of the spiral duct
%with $\Rstart$ and $\Rend$ as the bend radius at the start and end of the spiral duct,
and $\Nturns$ represents the number of turns of the spiral. The cross-section of the channel is uniform and has a square geometry with side length $l$. The horizontal and vertical co-ordinates within the square cross-section are denoted by $r$ and $z$, respectively, with the origin at the center of the square i.e. the domain is $-l/2\leq r \leq l/2$, $-l/2\leq z \leq l/2$. These cross-sectional co-ordinates are related to the global co-ordinates of the three-dimensional spiral duct as follows:
\begin{equation*}
    \mathbf{x}(\theta,r,z) = (R(\theta)+r)\cos(\theta) \,{\mathbf{i}}+(R(\theta)+r)\sin(\theta)\,{\mathbf{j}}+z\,{\mathbf{k}}.
\end{equation*}
The location of the particle's center is given by $\mathbf{x}_p=\mathbf{x}(\theta_p,r_p,z_p)$. In the absence of the particle, the incompressible steady fluid flow in a curved duct driven by a steady pressure gradient is referred to as Dean flow~\cite{Dean1927,Dean1959,Yamamoto_2004}. The presence of a particle disturbs this background Dean flow and the particle responds to this disturbed flow. Harding et al.~\cite{harding_stokes_bertozzi_2019} developed a general model for the leading order forces that govern the motion of such a particle in flow through a curved duct (constant $R(\theta)$) at sufficiently small flow rates. The forces on the particle from the fluid are calculated and used to construct a first order model for the trajectory of the particle giving the following dynamical equations of motion~\citep{harding_stokes_bertozzi_2019}:
\begin{equation*}
    \frac{\text{d} r_p}{\text{d} t}=-\text{Re}_p\frac{F_{s,r}}{C_r},\:\:\:
    \frac{\text{d} z_p}{\text{d} t}=-\text{Re}_p\frac{F_{s,z}}{C_z}\:\:\:\text{and}\:\:\:\frac{\text{d} \theta_p}{\text{d} t}=\frac{\bar{{u}}_a}{R/a+r_p},
\end{equation*}
where $\bar{{u}}_a$ is the axial component of the background fluid flow velocity, $F_{s,r}=\mathbf{F}_s\cdot\mathbf{e}_r$ and $F_{s,z}=\mathbf{F}_s\cdot\mathbf{e}_z$ are the radial and the vertical components of the cross-sectional force, respectively, with corresponding drag coefficients $C_r$ and $C_z$ that vary with the particle's position in the cross-section. The particle Reynolds number is $\text{Re}_p=\text{Re}\,(a/l)^2$, {where} $\text{Re}=\rho U_m l/\mu$ is the channel Reynolds number with $U_m$ the characteristic axial velocity of the background fluid flow. Here, we use a quasistatic approximation for the background fluid flow and extend this particle dynamics model to investigate particle dynamics in spiral ducts. This approximation is reasonable for spiral ducts with slowly changing curvature where the flow locally does not differ significantly from Dean flow in a constant curvature duct with the same curvature~\citep{Harding2018}. Numerical implementation of this model involves using a finite element method to compute the forces acting on the particle (see Harding et al.~\cite{harding_stokes_bertozzi_2019} for more details). Once the forces are pre-computed at numerous points in the cross-section and for numerous system parameter values, interpolants of $C_r$, $C_z$, $F_{s,r}$, $F_{s,z}$ are constructed and the particle dynamics are then simulated using the MATLAB solver ode45. For simulations of particle dynamics in a spiral duct, we fix the channel Reynolds number to $\text{Re}=25$. All the results presented herein are in dimensionless units with the dimensionless variables denoted by an overhead tilde and the lengths scaled by $l/2$.
   
 \begin{figure*}
\centering
\includegraphics[width=2\columnwidth]{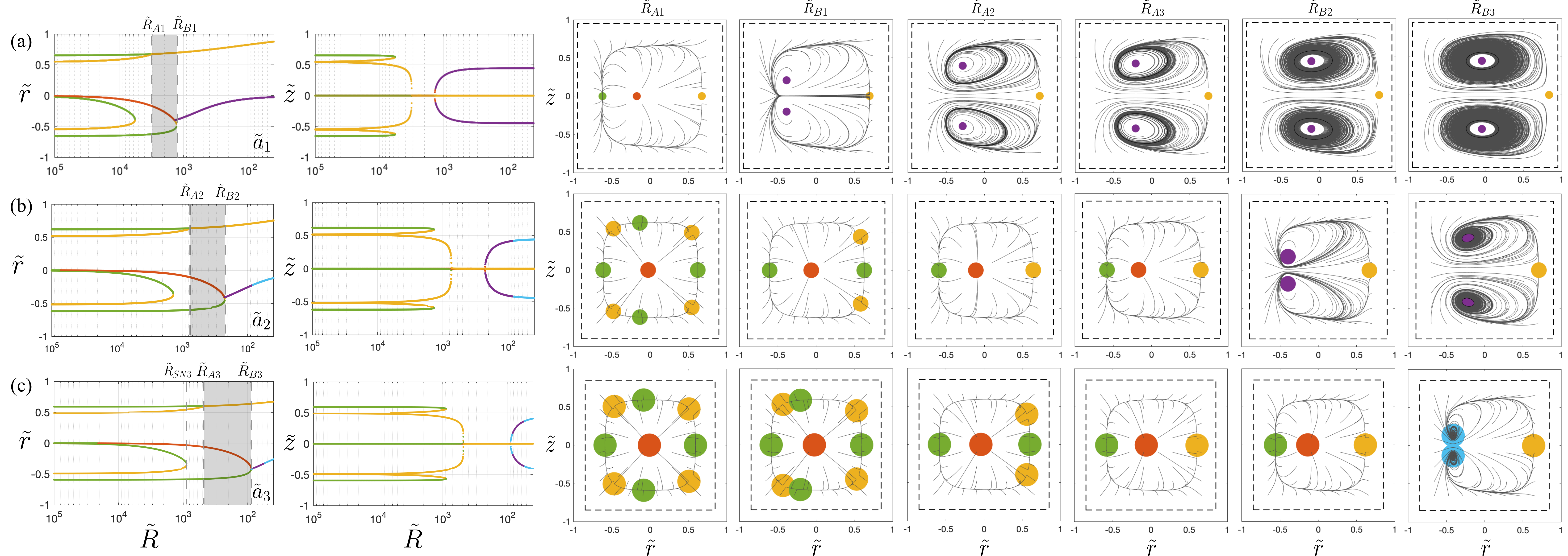}
\caption{Particle equilibria at different constant bend radii $\tilde{R}$ for three different particle sizes $\tilde{a}_1=0.05$, $\tilde{a}_2=0.10$ and $\tilde{a}_3=0.15$. The radial $\tilde{r}$ and vertical $\tilde{z}$ location of the equilibria are plotted as a function of the bend radius $\tilde{R}${; note that $\tilde R$ decreases from left to right}. The panels towards the right show the location of equilibria in the square cross-section as filled circles. The size of the circle corresponds to particle size and the color of the circles denotes the type of equilibria: unstable node in red, stable node in green, saddle point in yellow, unstable spiral in purple and a stable spiral in cyan. The parameters $\tilde{R}_{A1}\approx3000$, $\tilde{R}_{B1}\approx1250$, $\tilde{R}_{A2}\approx700$, $\tilde{R}_{B2}\approx210$, $\tilde{R}_{A3}\approx470$ and $\tilde{R}_{B3}\approx85$. The gray curves in each of these images illustrate typical trajectories of particles within the cross-section while the dashed square %represent the 
{shows} locations of the center of the particle for which %the particle
{it} will hit the walls of the duct. The gray shaded region in the left panels of (a), (b) and (c) corresponds to regions of a single stable node (SSN).}
\label{Fig: Particle equilibria}
\end{figure*}

Consider particles of three different sizes $\tilde{a}_1=0.05$, $\tilde{a}_2=0.10$ and $\tilde{a}_3=0.15$. For these particle sizes in a curved duct of constant bend radius, nine different particle equilibria exist inside the square cross-section for large bend radius (see Fig.~\ref{Fig: Particle equilibria} and also Fig.~2(a,b) of \citet{ValaniDSTA2021}); four stable nodes (green) near the center of the four edges, four saddle points (yellow) near the corners and an unstable node (red) near the center of the square cross-section. A slow manifold is formed along a closed curve that connects all the stable nodes and saddle points due to a large disparity in the magnitude of the two eigenvalues for these equilibria. 
%For each of these equilibria along the slow manifold, the eigenvalue in a direction tangential to the slow manifold is much smaller than the eigenvalue in a normal direction, resulting in slow migration along this manifold. 
As the bend radius is decreased progressively, a number of bifurcations take place. Firstly, saddle-node bifurcations take place where the stable nodes near the center of the top and bottom walls of the square collide and annihilate with the saddle points located near the inner wall. Further decreasing the bend radius results in the stable node located near the outer wall undergoing a subcritical pitchfork bifurcation with the two saddle points located above and below, and the three equilibria merge into a single saddle point. As the bend radius is yet further reduced, the unstable node near the center of the duct migrates towards the stable node located near the inner wall and undergoes a series of bifurcations~\citep{Valani2022SIADS}. Finally, at small bend radius, we get (i) unstable spirals with an encompassing limit cycle around them for a small particle of $\tilde{a}_1=0.05$, and (ii) stable spirals for $\tilde{a}_2=0.10$ and $\tilde{a}_3=0.15$. 

\begin{figure*}
\centering
\includegraphics[width=2\columnwidth]{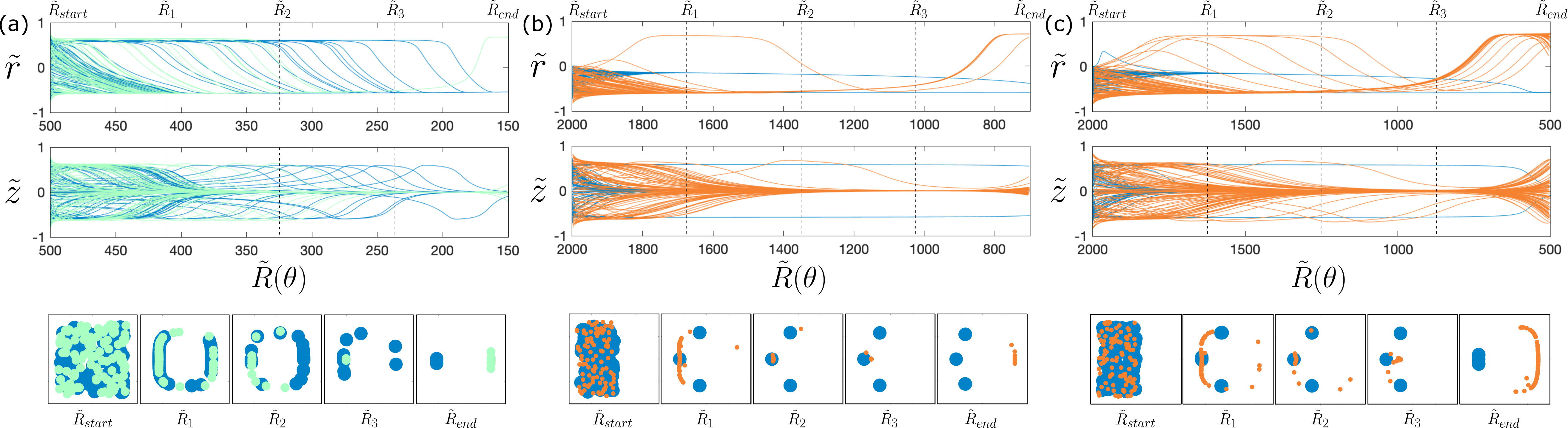}
\caption{Novel particle separation mechanism by exploiting bifurcations. Particle trajectories and snapshots of the cross-section at five equally spaced points along the spiral $\tRstart, \tilde{R}_1, \tilde{R}_2, \tilde{R}_3$ and $\tRend$, for a collection of non-interacting particles of two different sizes. (a) Particles of size $\tilde{a}_2=0.10$ (blue) and $\tilde{a}_3=0.15$ (green) with initial conditions randomly distributed over the entire cross-section and $\tRstart=500$ and $\tRend=150$ with $6$ turns, and particles of size $\tilde{a}_1=0.05$ (orange) and $\tilde{a}_3=0.15$ (green) with initial conditions randomly distributed over the inner half of the cross-section with spiral radii (b) $\tRstart=2000$ and $\tRend=700$ with $3$ turns, and (c) $\tRstart=2000$ and $\tRend=500$ with $2.5$ turns. See also supplemental videos S1 to S3.}
\label{Fig: Novel focusing}
\end{figure*}

We now demonstrate how these bifurcations can be utilized to separate two particles with small and large differences in size in an Archimedean spiral channel. We start by noting the bend radii corresponding to the existence of a single stable node (SSN) near the inner wall. This is shown by the gray shaded region in the left panels of Fig.~\ref{Fig: Particle equilibria}. For each particle size $\tilde{a}_i$, we denote the bend radius corresponding to start and end of the SSN region by $\tilde{R}_{Ai}$ and $\tilde{R}_{Bi}$ respectively, with $\tilde{R}_{Ai}>\tilde{R}_{Bi}$. We notice that with $\tilde{a}_3>\tilde{a}_2>\tilde{a}_1$, we have $\tilde{R}_{A3}<\tilde{R}_{A2}<\tilde{R}_{A1}$ and $\tilde{R}_{B3}<\tilde{R}_{B2}<\tilde{R}_{B1}$. Now, we classify the particles to be separated into two classes based on their SSN regions: (i) particle sizes for which their SSN regions overlap and (ii) particle sizes for which their SSN regions do not overlap.

Consider particles of size $\tilde{a}_2=0.10$ and $\tilde{a}_3=0.15$ as an example of the first class of particles. If we choose a spiral duct with $\tRstart$ near $\tilde{R}_{A3}$ and $\tRend\in[\tilde{R}_{B3},\tilde{R}_{B2}]$, then this selection will ensure that both particle sizes have a SSN near the inner wall at the start of the spiral. At the end of the spiral, particles of size $\tilde{a}_2$ will not have a SSN since $\tRend<\tilde{R}_{B2}$ while particles of size $\tilde{a}_3$ will still have a SSN. An example of particle dynamics with this selection of parameters is shown in Fig.~\ref{Fig: Novel focusing}(a) and supplemental video S1. Particles of both sizes initially snap onto the slow manifold and focus to their respective SSNs. As the particles flow along the spiral duct with decreasing bend radius and when $\tilde{R}(\theta)<\tilde{R}_{B2}$, the instability of the SSN of the smaller particles facilitates migration of focused $\tilde{a}_2$ sized particles along the center line (in the $\tilde{r}$ direction) of the square duct towards the saddle point located near the outer wall. Due to a large disparity in the two eigenvalues of this saddle point, the particle will spend a long time near this saddle point before being ejected vertically in the $\tilde{z}$ direction. Meanwhile, the bigger $\tilde{a}_3$ particles remain at the SSN near the inner wall at the end of the spiral. Hence, we get the bigger particles focused near the inner wall and the smaller particles transiently focused near the outer wall. If the spiral is terminated before the $\tilde{a}_2$ sized particles start leaving the saddle point, then one can separate the two particle sizes at opposite ends of the cross-section, thus achieving a high separation resolution. We note that a separation mechanism similar to the one described here might be at play in the recent experimental work of \citet{Cruz2021} where they demonstrated separation of two similar sized sub-micron particles in a specific High Aspect Ratio Curved (HARC) rectangular duct comprised of two stages. However, we believe that the separation mechanism outlined above may work more broadly for various cross-sections and curved duct geometries to separate two particles when their SSN regions overlap along with the corresponding bifurcations. %Moreover, in principle, this mechanism can potentially be applied to separated two closely spaced particle sizes as well (see Appendix~\ref{app: close sizes} for an example).  

For the second class of particles having no overlap between their SSN regions, for example, particles of size $\tilde{a}_1=0.05$ and $\tilde{a}_3=0.15$, we cannot use the above described mechanism to separate the particles because multiple stable nodes exist for $\tilde{a}_3$ for the bend radii corresponding to the SSN region of $\tilde{a}_1$. However, if we inject particles only on half of the cross-section, rather than allowing particles to be anywhere in the cross-section, then we can focus the larger particles selectively to specific stable nodes leading to a clean separation between the two particle sizes. For example, we choose the start bend radius $\tRstart$ of the spiral to be near $\tilde{R}_{A1}$ and the ending radius to be such that $\tRend<\tilde{R}_{B1}$, and initiate the particles in the inner half of the cross-section, as is commonly done in DFF. With this choice of parameters the smaller particles initially focus to the SSN near the inner wall and then migrate towards the outer wall once the SSN becomes unstable. The bigger particles will focus to either (i) the three stable nodes near the inner wall if $\tRend>\tilde{R}_{SN3}$ where $\tilde{R}_{SN3}$ is the bend radius corresponding to the saddle-node bifurcations for $\tilde{a}_3$ near the inner wall (see Fig.~\ref{Fig: Novel focusing}(b) and supplemental video S2), or (ii) a SSN near the inner wall if $\tRend<\tilde{R}_{SN3}$ (see Fig.~\ref{Fig: Novel focusing}(c) and supplemental video S3). In either case, we will achieve a large separation in the $\tilde{r}$ direction between these two different sized particles. 
We note that we can also implement starting the particles on the inner half of the duct cross-section for the first class of particle shown in Fig.~\ref{Fig: Novel focusing}(a) and this may result in quicker particle separation.

%\begin{itemize}
%    \item Say that Fig.~\ref{Fig: Fig2} shows an example of separation of two particles of size $a_1$ and $a_2$ achieved by exploiting the aforementioned bifurcation
%    \item Here the bigger particle focuses near the inner wall after travelling along the slow manifold 
%    \item The smaller particle initially prefocused to the inner wall as well but after that fixed point becomes unstable, this particle travels along the centerline to the saddle point where it stay there for a long time due to the small magnitude eigenvalues for the outgoing trajectory
%    \item Show that in this way we can separate two particle by exploiting the bifurcations that take place in a spiral duct
%\end{itemize}

\begin{figure}
\centering
\includegraphics[width=\columnwidth]{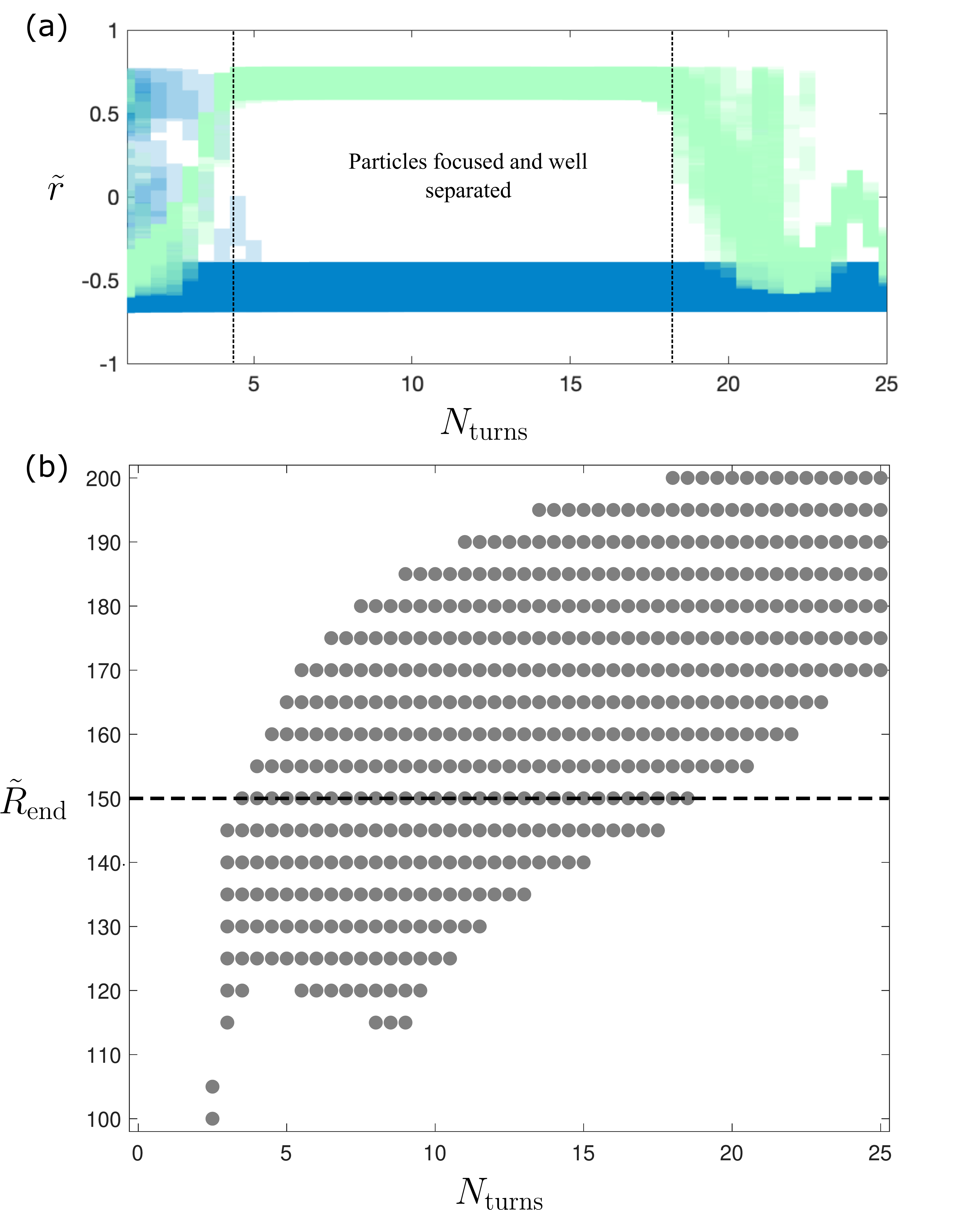}
\caption{Particle separation with variations in the number of turns $\Nturns$ and the radius $\tRend$ at the end of the spiral duct, with fixed $\tRstart=500$. (a) Horizontal position $\tilde{r}$ of particles of size $\tilde{a}_2=0.10$ (green) and $\tilde{a}_3=0.15$ (blue) at the end of the duct as a function of the number of turns for an Archimedean spiral with $\tRend=150$. (b) Regions of particle separation (gray) in the parameter space formed by $\Nturns$ and $\tRend$. The white region corresponds to particles either not focused in the horizontal direction (when the horizontal standard deviation of either cluster is more than the corresponding particle radius) or not well separated (when the horizontal separation between the cluster centers is less than $3(\tilde{a}_2 + \tilde{a}_3)$). The horizontal dashed black line corresponds to the plot in panel (a) for $\tRend=150$.}
\label{Fig: Fig4}
\end{figure}

We now explore the influence of the number of turns of the spiral duct on the focusing behavior by taking the particles $\tilde{a}_2$ and $\tilde{a}_3$ from the first class. Figure~\ref{Fig: Fig4}(a) shows the radial $\tilde{r}$ location of particles $\tilde{a}_2$ and $\tilde{a}_3$ at the end of the spiral duct $\tRend$ as a function of the number of turns $\Nturns$. We see that for $\Nturns\lesssim 4$, the particles are not well separated. Due to the shorter length of the spiral, the particles cannot sufficiently focus to the desired equilibria, resulting in poor particle focusing and separation. For $5 \lesssim \Nturns \lesssim 18$, the particles are focused and well separated at the two ends of the duct in the $\tilde{r}$ direction. For $\Nturns\gtrsim 18$, the separation again becomes poor as the smaller particles start diverging from the saddle point along the slow manifold and eventually migrate to a limit cycle around the unstable spiral near the inner wall. Hence, we find a range of turns ($\Nturns$) for effective particle separation. To further identify regions of optimal separation, Fig.~\ref{Fig: Fig4}(b) shows regions in the parameter space formed by $\Nturns$ and $\tRend$ where the particles are focused (the horizontal standard deviation of each cluster is less than the corresponding particle radius) and well separated (the horizontal separation between the cluster centers is more than three times the minimum separation distance of $\tilde{a}_2 + \tilde{a}_3$). 
%In this parameter space, we identify a band corresponding to a large particle separation in the horizontal direction between the $\tilde{a}_2$ and $\tilde{a}_3$ sized particles. Moreover, we observe a periodic structure in the separation distance for small $\tRend$ and large $\Nturns$. This due to oscillations in the distance between the centers of two clusters as the smaller particles approach a stable spiral after leaving the saddle-point. 
Parameter-space plots such as Fig.~\ref{Fig: Fig4} may aid in the design of spiral microfluidic devices that employ this mechanism to separate particles. 

%\begin{itemize}
%    \item To see the No of turns of the spiral geometry required to separate particles using this mechanism, we plot the final location of the two different sized particles at the end of the spiral
%    \item This shows that when the number of turns of the particles are in the range between $N_1$ and $N_2$ then can be separated well using this mechanism. For too low number of turns the particles don't get enough time to focus to the equilibrium points
%    \item If the number of turns are too high then the smaller particle will leave the saddle point and travel towards either a stable spiral or a limit cycle encompassing an unstable spiral.
%\end{itemize}

%\begin{itemize}
%    \item I can potentially also include another plot where I can fixed the size of one of the particles say $a_1$ and then keep the number of turns fixed and vary $a_2$ to see how close the particle sizes I can get which can be separated using this method. However, this might also require tuning of the starting and the ending radius of the spiral. So rather it might be better to take two closely spaced particle sizes and show that they can be separated using this mechanism for a given number of turns.
%    \item It would be good to somehow show the range of particle sizes that can be separated using this mechanism and then that can eventually show that two similar particle sizes can also be separated using this mechanism. 
%\end{itemize}

%\section{Conclusions}

We have shown how bifurcations in the particle equilibria with respect to duct bend radius can be used to separate particles with both small and large size differences in spiral channels. 
%This mechanism provides a novel way to separate particles in inertial microfluidics by utilizing bifurcations in particle equilibria. 
In inertial microfluidic experiments, the presence of experimental noise may alter the stability properties of the particle equilibria which may hinder or enhance the ability to separate particles using this mechanism. However, the observation of a similar mechanism working in the experiments of \citet{Cruz2021} is promising. Nevertheless, it would be fruitful to perform systematic experiments to determine the extent to which this separation mechanism can be utilized to separate particles. Lastly, although we have restricted ourselves to certain particle sizes and duct geometries to demonstrate the mechanism, the general idea of using bifurcations in particle equilibria to facilitate transient particle separation may be applied more broadly to different duct geometries and particle sizes.

% The \nocite command causes all entries in a bibliography to be printed out
% whether or not they are actually referenced in the text. This is appropriate
% for the sample file to show the different styles of references, but authors
% most likely will not want to use it.
%\nocite{*}

\begin{acknowledgments}
This research is supported under Australian Research
Council’s Discovery Projects funding scheme (project number DP200100834). The results were computed using supercomputing resources provided by the Phoenix HPC service at the University of Adelaide and the Raapoi
HPC service at Victoria University of Wellington.
\end{acknowledgments}

\section*{Conflict of Interest}
The authors have no conflicts to disclose.

\section*{Data Availability Statement}
The findings of this study are based on the model developed by \citet{harding_stokes_bertozzi_2019} and the corresponding data available at \url{https://github.com/brendanharding/ILFHC}. Further data supporting the findings of this study is available from the corresponding author upon reasonable request.

% Create the reference section using BibTeX:
\bibliography{aipsamp}

\end{document}